\documentclass{PoS}
\usepackage{dsfont}
\usepackage{amsmath}
\usepackage{amsfonts}
\usepackage{amssymb}

\title{Features of the Refined Gribov-Zwanziger theory: propagators, BRST soft symmetry breaking and glueball masses}

\ShortTitle{Features of the Gribov-Zwanziger theory and of its refined version}

\author{\speaker{Silvio P. Sorella}\thanks{Work supported by FAPERJ, Funda{\c c}{\~a}o de Amparo {\`a} Pesquisa do Estado do Rio de Janeiro, under the program {\it Cientista do Nosso Estado}, E-26/101.578/2010}\\
        Departamento de F\'{\i }sica Te\'{o}rica, Instituto de F\'{\i }sica, UERJ - Universidade do Estado do Rio de Janeiro, Rua S\~{a}o Francisco Xavier 524, 20550-013 Maracan\~{a}, Rio de Janeiro, Brasil\\
        E-mail: \email{sorella@uerj.br}}

\author{David Dudal\\
        Ghent University, Department of Physics and Astronomy, Krijgslaan 281-S9, 9000 Gent, Belgium\\
        E-mail: \email{david.dudal@ugent.be}}

\author{Marcelo S. Guimar\~aes\\
        Departamento de F\'{\i }sica Te\'{o}rica, Instituto de F\'{\i }sica, UERJ - Universidade do Estado do Rio de Janeiro, Rua S\~{a}o Francisco Xavier 524, 20550-013 Maracan\~{a}, Rio de Janeiro, Brasil\\
        E-mail: \email{msguimaraes@uerj.br}}

\author{Nele Vandersickel\\
        Ghent University, Department of Physics and Astronomy, Krijgslaan 281-S9, 9000 Gent, Belgium\\
        E-mail: \email{nele.vandersickel@ugent.be}}

\abstract{The present work discusses an approach to access the physical spectrum of the Yang-Mills theory quantized in the Landau gauge. By using recent lattice data on the gluon propagator,  it is possible to study the two-point functions of gauge invariant composite operators, from which masses of glueballs can be extracted. It turns out that the momentum dependence of the gluon propagator is very well reproduced by the corresponding tree-level gluon propagator obtained from the Refined Gribov-Zwanziger theory, which takes into account the presence of the Gribov horizon as well as the effect of condensates of mass dimension two. The resulting glueball masses are in good agreement with the available lattice data.}

\FullConference{The many faces of QCD\\
		November 1-5, 2010\\
		Gent Belgium}

\begin{document}

\section{Introduction}

\noindent The characterization  of the physical spectrum of QCD at low energy scales from the first principles of the theory is an outstanding problem. QCD is a strongly interacting theory at low energy scales and thus the usual perturbative treatment does not apply. Due to color confinement, there is no free particle interpretation for the elementary fields appearing in the Lagrangian formulation of the theory. As a consequence, the properties of the physical spectrum are expected to be encoded in the correlation functions of suitable composite operators built with the elementary fields. \\\\ The numerical lattice studies of the Yang-Mills sector of QCD, with the accurate determination of  the ghost and gluon propagators, have provided   valuable results. Gluons show unphysical properties at low energies with a clear violation of positivity of the propagator. Although this behavior is expected in a confining theory, the employment of a positivity violating gluon propagator for a direct analytic evaluation of the correlation functions of the composite operators corresponding to the states of the spectrum is a highly nontrivial issue. \\\\In the following, we aim at showing that the decoupling type gluon propagator obtained from the Refined Gribov-Zwanziger theory yields a useful analytic set up for an estimate of the masses of the first glueball states with quantum numbers $J^{PC} = 0^{++}, 0^{-+}, 2^{++}$, while being in very good agreement with the recent lattice data.\\\\It is worth emphasizing that the framework we are going to outline here can, to some extent, be considered as model independent.  In fact,  the gluon propagator could be obtained by performing a good fit of the lattice data. The resulting expression could be employed to work out
the correlation function of the composite operators corresponding to the aforementioned glueball states. Evidently, this is a model independent set up.  It is remarkable thus that the gluon propagator obtained from the Refined Gribov-Zwanziger gives a very good fit of the lattice data in the non-perturbative low energy region, while providing a workable  analytic structure.
\section{A short survey on the Gribov-Zwanziger theory}

\noindent The Yang-Mills theory is a gauge theory, thus, in order to be consistently quantized the gauge redundancy has to be removed. However,  imposing a local, Lorentz covariant, condition which would select one and only one representative for each gauge orbit is not possible \cite{Singer:1978dk}.\\\\The problem was first noticed by Gribov \cite{Gribov:1977wm} who observed that the Landau gauge condition $\partial_{\mu} A^a_{\mu} = 0$  is not sufficient to account for the Gribov copies, {\it i.e.} gauge transformed configurations which still obey the Landau condition.
Consider, for example, an infinitesimal gauge transformation $A^a_{\mu} \rightarrow  A^a_{\mu} + D_{\mu}^{ab}\omega^b$. The gauge transformed field would fulfill the Landau condition if  $\partial_{\mu}D_{\mu}^{ab}\omega^b = 0$. Zero modes of the Faddeev-operator ${\cal M}^{ab}$
\begin{align}
{\cal M}^{ab}=-\partial_\mu D_\mu^{ab},\;\; D_\mu^{ab}=\partial_\mu\delta^{ab}-gf^{abc}A_\mu^c\;, \label{fpoperator}
\end{align}
provide thus examples of Gribov copies. In order to take into account the existence of the gauge copies,  Gribov \cite{Gribov:1977wm} proposed to restrict the domain of integration in the functional integration to the so called Gribov region $\Omega$.\footnote{It is known that additional gauge copies still remain within the Gribov region. The elimination of all copies is, by definition, only attained by restricting the domain of integration to the \emph{Fundamental Modular Region}. However, till now, a way to  implement the restriction to the FMR within a local field theory framework is not available.}:
\begin{align}
\Omega=\left\{A_\mu^a|\partial_\mu A_\mu^a=0, {\cal M}^{ab}>0\right\}\;. \label{gribovregion}
\end{align}
The boundary of this region is known as the Gribov horizon and is the locus where the lowest eigenvalue of ${\cal M}$ vanishes. Later on, Zwanziger \cite{Zwanziger:1989mf} was able to show that the restriction to the region $\Omega$ can be done within a local and renormalizable field theory framework. The resulting action  is known as the Gribov-Zwanziger action
\begin{eqnarray}
S_{GZ} & = & \int d^4 x \left(  \frac1{4} F_{\mu\nu}^{a}F_{\mu\nu}^{a} + ib^a\partial_{\mu}A^a_\mu + {\bar c}^a \partial_\mu D_{\mu}^{ab} c^b  \right) \nonumber \\
{\ }{\ }{\ } & + &\int d^4x \left( -  {\bar \varphi}^{ac}_\mu \partial_\nu D_{\nu}^{ab} \varphi^{bc}_{\mu} +  {\bar \omega}^{ac}_\mu \partial_\nu D_{\nu}^{ab} \omega^{bc}_{\mu} + g f^{amb} (\partial_{\nu}{\bar \omega}^{ac}_{\mu}) (D^{mp}_{\nu} c^p)\varphi^{bc}_{\mu} \right) \nonumber \\
{\ }{\ }{\ } & +&   \int d^4x\left(\gamma^2\,g\,f^{abc}A_\mu^{a}(\varphi_\mu^{bc}-\bar{\varphi}_\mu^{bc})-d(N^2-1)\gamma^4 \right) \;, \label{GZact}
\end{eqnarray}
where $d=4$ stands for the space-time dimensions. The dimensionful parameter  $\gamma$ is known as the Gribov parameter.
This expression can be written as
\begin{eqnarray}
S_{GZ}  = \frac1{4}\int d^4 x \;F_{\mu\nu}^{a}F_{\mu\nu}^{a} + s\int d^4 x \left(\bar{c}^a\partial_{\mu} A^a_{\mu} - \bar{\omega}_\mu^{ac}\partial_\nu D^{ab}_\nu \varphi_{\mu}^{bc}\right) + S_{\gamma} \;, \label{GZact}
\end{eqnarray}
with $S_\gamma$ given by
\begin{align}
S_{\gamma}&=\int d^4 x\left(\gamma^2\,g\,f^{abc}A_\mu^{a}(\varphi_\mu^{bc}-\bar{\varphi}_\mu^{bc})-d(N^2-1)\gamma^4 \right) \;, \label{BRSbreak}
\end{align}
and where $s$ denotes for the nilpotent BRST operator, whose action on the fields is specified by
\begin{align}
sA_\mu^{a} &= - D^{ab}_\mu c^b = -( \partial_\mu \delta^{ab} +  g f^{acb} A^c_\mu)c^b\;,\nonumber \\
sc^a &=  \frac{g}{2} f^{acb}c^b c^c\;,\nonumber\\
s\bar{c}^a &= ib^a\;,\qquad sb^a=0\;,\nonumber\\
s\bar{\omega}_\mu^{ab} &= \bar{\varphi}_\mu^{ab}\;,\qquad s\bar{\varphi}_\mu^{ab}=0\;,\nonumber\\
s\varphi_\mu^{ab} &= \omega_\mu^{ab}\;,\qquad s\omega_\mu^{ab}=0\;. \label{BRS}
\end{align}
The fields $\{\varphi_\mu^{ab},\overline\varphi_\mu^{ab}\}$ and $\{\omega_\mu^{ab}, \overline\omega_\mu^{ab}\}$ are a set of  bosonic, resp.~fermionic fields introduced in order to express the action in  local form. When the Gribov horizon is removed, which amounts to formally set $\gamma =0$, these fields give rise to a BRST quartet. They decouple from the theory, and the Gribov-Zwanziger action reduces to the ordinary Faddeev-Popov action.  However, one has to observe that the  parameter $\gamma$ is not free, being determined in a self consistent way by a gap equation, called the horizon condition \cite{Zwanziger:1989mf}. It reads
\begin{align}
\frac{\partial {\cal E}_{vac}}{\partial \gamma^2}=0 \;,  \label{geq}
\end{align}
where $ {\cal E}_{vac} $ is the vacuum energy
\begin{align}
e^{- {\cal E}_{vac} } = \int [d\Phi] \; e^{-S_{GZ}} \;,  \label{vc}
\end{align}
and $[d\Phi]$ stands for the functional integration over all fields appearing in $S_{GZ}$.\\\\An important property of the action $S_{GZ}$ is its multiplicative renormalizability \cite{Zwanziger:1989mf,Maggiore:1993wq,Gracey:2006dr,Dudal:2008sp}
. This is a highly non-trivial feature that, among other things, enables us to consistently introduce into the theory composite operators and establish their renormalization properties.
\section{The issue of the BRST symmetry}

\noindent In the absence of the term $S_{\gamma}$, the action \eqref{GZact} enjoys BRST invariance. In fact
\begin{equation}
s   \int d^4 x \left(\frac1{4} F_{\mu\nu}^{a}F_{\mu\nu}^{a} + s  \left(\bar{c}^a\partial_{\mu} A^a_{\mu} - \bar{\omega}_\mu^{ac}\partial_\nu D^{ab}_\nu \varphi_{\mu}^{bc}\right)     \right) =0  \;.
\end{equation}
The Gribov-Zwanziger action is, however, not left invariant by the BRST transformations, eqs.\eqref{BRS}, which are broken by the term $S_{\gamma}$, namely
\begin{align}
sS_{GZ} = sS_{\gamma} = \gamma^2\,\int d^4x\left(-g\,f^{abc}(D^{ad}_\mu c^d) (\varphi_\mu^{bc}-\bar{\varphi}_\mu^{bc}) +g\,f^{abc}A_\mu^{a}\omega_\mu^{bc}\right)\;. \label{BRSactbreak}
\end{align}
Though, the breaking of the BRST symmetry in the GZ framework is of a very special nature. Notice that the breaking term, being of dimension two in the fields, is a soft breaking. This fact ensures the renormalizability of the theory through suitable Ward identities \cite{Zwanziger:1989mf,Maggiore:1993wq,Gracey:2006dr,Dudal:2008sp}. Recently, there have been several developments in the understanding of this breaking.  Notice that the breaking is quadratic in the fields. As such,  it has to be treated as a composite field operator, a feature which requires the introduction of a suitable set of external sources in order to implement the Slavnov-Taylor identities. In \cite{Capri:2010hb},  it has been shown that this breaking can be converted into a linear one. As a consequence, the resulting nilpotent linearly broken BRST symmetry can be directly employed to obtain a system of Slavnov-Taylor identities. In \cite{Sorella:2009vt,Kondo:2009qz}, it was pointed out that the softly broken BRST symmetry of the GZ action can be converted into a exact symmetry, however non-local. This non-local invariance has been localized in \cite{Dudal:2010hj}, though the resulting BRST symmetry is not nilpotent. Also, there are attempts to investigate the soft BRST breaking as a spontaneous symmetry breaking \cite{Zwanziger:2010iz}.\\\\In summary, we can safely state that these results have already provided a certain understanding of the important issue of the BRST symmetry versus the Gribov horizon.  The main result obtained so far remains that of the renormalizability of the theory. As a byproduct,  let us mention the very important property that a local, gauge invariant operator ${\cal O}$ can be promoted to a renormalized operator ${\cal O}_R$, so that the correlation functions
\begin{align}
 \langle {\cal O}_R(k) {\cal O}_R(-k) \rangle \label{corrfunc}
\end{align}
can be consistently evaluated order by order within the GZ framework \cite{Dudal:2009zh}.

\section{Dimension two condensates and the Refined Gribov-Zwanziger theory}

\noindent The action (\ref{GZact}) is supposed to account for nonperturbative infrared features of Yang-Mills theory. The development of a non-zero value for the Gribov parameter $\gamma$ from the gap equation \eqref{geq} is a manifestation of the nontrivial properties of the vacuum of the theory induced by restriction to the Gribov horizon. Moreover, there are additional sources on nonperturbative effects which can be encoded in a set of dimension 2 condensates, namely $\langle A^a_{\mu}A^a_{\mu} \rangle$ and $\langle \overline\varphi_\mu^{ab} \varphi_\mu^{ab}-\overline\omega_\mu^{ab}\omega_\mu^{ab}\rangle$. It is possible to explicitly take into account the contribution of these condensates by introducing the corresponding dimension two operators directly into the GZ action. The resulting formulation is, by now, known as the Refined Gribov-Zwanziger theory \cite{Dudal:2008sp}, and amounts to the following modification of the original Gribov-Zwanziger theory:
\begin{align}
 S_{GZ} \rightarrow S_{RGZ}=S_{GZ}+\int d^4x \left(\frac{m^2}{2} {A^a_{\mu}}{A^a_{\mu}} - {M^2} \left(\overline\varphi_\mu^{ab} \varphi_\mu^{ab}-\overline\omega_\mu^{ab}\omega_\mu^{ab}\right)\right) \;,  \label{RGZ}
\end{align}
As well as the Gribov-Zwanziger action, the refined action $S_{RGZ}$ also enjoys multiplicative renormalizability \cite{Dudal:2008sp}.

\section{A look at the gluon and ghost propagators}

\noindent Let us give a look at the gluon and ghost propagators obtained from both GZ and RGZ actions, respectively.  Let us start with the GZ formulation. Here, the tree-level gluon propagator in $d=4, 3$ and $2$ dimensions is given by
\begin{align}
\langle A^a_{\mu}(k)A^b_{\nu}(-k) \rangle_{GZ} = \delta^{ab}\left(\delta_{\mu\nu} - \frac{k_{\mu}k_{\mu}}{k^2} \right)\frac{k^2}{k^4+\lambda^4}\;,  \label{GZgluonprop}
\end{align}
where $\lambda^4 = 2g^2 N\gamma^4$. The gluon propagator turns out to be suppressed in the infrared region, attaining a vanishing value at zero momentum, $k=0$. On the other hand, the ghost propagator displays an enhanced behavior in the infrared
\begin{align}
\langle \bar{c}^a(k)c^b(-k) \rangle_{GZ}  \sim \delta^{ab} \frac{1}{k^4}\;, \;\;\; \text{for} \;\; k^2 \sim 0.\label{GZghostprop}
\end{align}
A gluon suppression and ghost enhancement are only observed for lattice data in 2 dimensions. They are not in agreement with the most recent lattice data for $d=4,3$  dimensions \cite{Cucchieri:2007md,Cucchieri:2007rg,Cucchieri:2009zt}, which  seem to point towards a suppressed gluon propagator which attains a non-vanishing value at $k=0$, and  to a non-enhanced ghost at $k\sim0$ which keeps essentially the free behavior $\sim \frac{1}{k^2}\Big|_{k\sim0}$. \\\\On the other hand, the gluon propagator obtained from the RGZ formulation in $d=4,3$ dimensions shows a different behavior. As in the GZ case, the propagator is suppressed in the infrared. However, differently from the GZ case, it attains now a nonvanishing value at $k=0$, as one observes from
\begin{align}
\langle A^a_{\mu}(k)A^b_{\nu}(-k) \rangle_{RGZ} = \delta^{ab}\left(\delta_{\mu\nu} - \frac{k_{\mu}k_{\mu}}{k^2} \right)\frac{k^2+M^2}{k^4 + (M^2+m^2)k^2 +\lambda'^4}\;,  \label{RGZgluonprop}
\end{align}
where $\lambda'^4 = 2g^2 N\gamma^4 + m^2 M^2$. This expression is in very good agreement with the numerical  data.
In fact, in $d=4$, an accurate fit with the SU(3) lattice data is possible for momentum scales up to $k \sim 1.5$ GeV \cite{Dudal:2010tf}. The fit is established for the following values of parameters
\begin{align}
M^2+m^2\approx 0.337 \text{ GeV}^2\,\,, M^2\approx 2.15 \text{ GeV}^2\,\,, \lambda'^4\approx0.26\text{ GeV}^4\,. \label{RGZmasses}
\end{align}
\begin{figure}[h]
       \centering
       \includegraphics[scale=0.3]{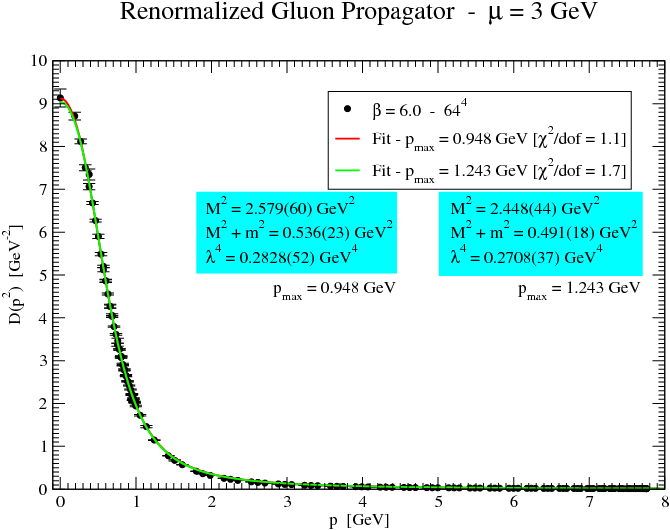}
       \includegraphics[scale=0.3]{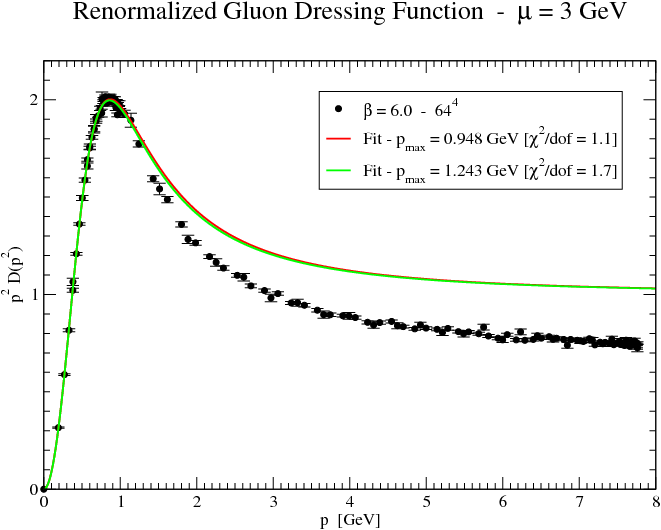}
       \caption{Fit to the gluon propagator (left) and dressing function (right). Figures taken from \cite{Dudal:2010tf}. \label{fig:3}}
\end{figure}

\noindent Also, the ghost propagator obtained from RGZ in both $d=4,3$ displays a behavior in harmony with the lattice data
\begin{align}
\langle \bar{c}^a(k)c^b(-k) \rangle_{RGZ}  \sim \delta^{ab} \frac{1}{k^2}\;, \;\;\; \text{for} \;\; k^2 \sim 0.\label{RGZghostprop}
\end{align}
We also point out that in $d=2$ the RGZ and the GZ theories coalesce, due to severe infrared divergences which do not allow to consistently introduce dimension two condensates. Due to this property, in $d=2$ the RGZ theory gives the same results of the GZ theory. Therefore,  the RGZ gives rise to gluon and ghost propagators which are in good agreement with the lattice data in $d=4,3,2$ dimensions. This is a remarkable achievement.

\section{$i$-particles and the glueball spectrum}

\noindent Both GZ and RGZ gluon propagators are associated to unphysical excitations.  In fact, it is straightforward to verify that the propagators (\ref{GZgluonprop}) and (\ref{RGZgluonprop}) each have two complex conjugated poles. Let us concentrate on the RGZ propagator.  In this case, from (\ref{RGZmasses}),  we can deduce the numerical values of the complex conjugated masses which, in GeV$^2$ units, are
\begin{align}
m^2_{\pm} = \mu^2 \pm \sqrt{2} \theta^2 = 0.1685 \pm 0.4812i.\label{ipmasses}
\end{align}
The unphysical excitation associated with these complex masses have been called $i$-particles \cite{Baulieu:2009ha}. \\\\The $i$-particles diagonalize the quadratic part of the RGZ action, namely
\begin{eqnarray}
S_{RGZ} = \int d^4x  \left[ \frac{1}{2} {\lambda}^{a}_{\mu} \left( -\partial^2  + m_+^2 \right)  {\lambda}^{a}_{\mu} + \frac{1}{2} {\eta}^{a}_{\mu}\left( -\partial^2 + m_-^2  \right)   {\eta}^{a}_{\mu}  +\text{rest}\right]\,.\label{RGZipart}
\end{eqnarray}
They provide a useful set up in order to extract the analytic properties of correlation functions of  gauge invariant composite operators. For instance, the abelian part of the gauge invariant composite operator $F^a_{\mu\nu}F^a_{\mu\nu}$ can be decomposed in the following way:
\begin{equation}\label{scalaripart}
\left( F^a_{\mu\nu}F^a_{\mu\nu}\right)\Big|_{abelian} =\left( \partial_\mu A^a_\nu -\partial_\nu A^a_\mu \right)^2 =\lambda_{\mu\nu}^a\eta_{\mu\nu}^a+\text{rest}\,.
\end{equation}
where $\lambda_{\mu\nu}^a=\partial_\mu\lambda_\nu^a-\partial_\nu\lambda_\mu^a$ and $\eta_{\mu\nu}^a=\partial_\mu\lambda_\nu^a-\partial_\nu\lambda_\mu^a$ are the abelian field strengths of the $i$-particles fields. The first term,  $\lambda_{\mu\nu}^a\eta_{\mu\nu}^a$, of expression \eqref{scalaripart} can be shown to be associated to the part of the two-point correlation function of the operator $\left( F^a_{\mu\nu}F^a_{\mu\nu}\right)\Big|_{abelian} $ which displays a cut along the negative real axis in the complex Euclidean $k^2$ plane, thus having a physical interpretation, see \cite{Baulieu:2009ha} for details. The remaining part, called {\it rest} in eq.\eqref{scalaripart},  displays cuts along the imaginary axis and, as such, has no physical interpretation.  Notice that the operator $F^a_{\mu\nu}F^a_{\mu\nu}$ has the quantum numbers of the glueball state $J^{PC}=0^{++}$. In order to study the glueball spectrum,  in a lowest order approximation, we shall consider thus the abelian components of suitable composite operators  ${\cal O}(x)$ displaying the quantum numbers $J^{PC}$ and which, as in the case of $\left( F^a_{\mu\nu}F^a_{\mu\nu}\right)\Big|_{abelian} $, are built with pairs of $i$-particles.  Further,  we look at the two-point functions of the operators ${\cal O}(x)$. In order to have a physical meaning, these correlation functions should display a meaningful spectral representation, {\it i.e.}
\begin{align}
\langle {\cal O}(k) {\cal O}(-k) \rangle = \int_{0}^{\infty}dt \frac{\rho(t)}{t+k^2} \;. \label{KL}
\end{align}
with the spectral function $\rho(t)$ a positive definite quantity. The spectral function has the general form $\rho(t) = Z \sum_i \delta(t-m_i) + A\theta(t-t_0)$, whith $Z$ and $A$ positive quantities. The  scales $m_i$ correspond to the poles of the correlation functions, thus identifying the  masses of the  physical particles of the spectrum. Moreover,  $t_0$ is  the threshold for a multi-particles state. We see thus that the spectral function carries important information about the physical spectrum of the theory.

\noindent There are various methods to extract information on the spectrum from the knowledge of the spectral function, see, for example,  \cite{Narison:2002pw}. In the present case,  a modified reformulation of the moment problem \cite{Dudal:2010cd}, well adapted to the infrared region, was used to obtain estimates of the glueball masses of the scalar $0^{++}$, pseudoscalar $0^{-+}$ and tensor $2^{++}$ glueballs. The corresponding glueball operators in the $i$-particles representation are given by
\begin{eqnarray}
{\cal O}_{0^{++}}(x) & = &  \lambda^a_{\mu\nu}(x)\eta^a_{\mu\nu}(x) + {\ }\text{rest}, \label{io++} \\
\left[ {\cal O}_{2^{++}}(x) \right]_{\mu\nu} & = &  \left(P_{\mu\alpha}P_{\nu\beta} - \frac 13 P_{\mu\nu}P_{\alpha\beta}\right) \left(   \lambda^a_{\alpha\sigma}(x)\eta^a_{\beta\sigma}(x) + \eta^a_{\alpha\sigma}(x)\lambda^a_{\beta\sigma}(x)\right) + {\ }\text{rest},  {\ }{\ }{\ }{\ }{\ }{\ }\label{i2++}\\
{\cal O}_{0^{-+}}(x) & = & \frac{1}{2} \varepsilon_{\mu\nu\rho\sigma} \left( \lambda^a_{\mu\nu}(x) \eta^a_{\rho\sigma}(x)+ \eta^a_{\mu\nu}(x) \lambda^a_{\rho\sigma}(x)\right)  + {\ }\text{rest}. \label{io-+}
\end{eqnarray}
Using the lattice input for the $i$-particle masses (\ref{ipmasses}), the following estimates have been obtained:
\begin{equation}
  m_{0++}\approx 1.96 \text{ GeV}\;, \;\;\; m_{0-+}\approx 2.19 \text{ GeV}\;, \;\;\; m_{2++} \approx 2.04 \text{ GeV}\,. \label{gluemasses}
\end{equation}
Note that the observed mass hierarchy $m_{0++}<m_{2++}<m_{0-+}$ is reproduced\footnote{A similar hierarchy has also  been obtained in \cite{Capri:2010pg}, where the $i$-particles set up  has been combined with the SVZ sum rules.}. By comparing these numbers with the lattice results
\begin{equation}
m_{0++}^{\text{lat}}\approx 1.73 \text{ GeV}\,, m_{0-+}^{\text{lat}}\approx 2.59 \text{ GeV}\,, m_{2++}^{\text{lat}}\approx 2.40 \text{ GeV}\,, \label{gluemasseslattice}
\end{equation}
we can see a rather good agreement, all data being within a $20\%$ range of approximation.

\section{Conclusions}

Our main conclusion is that  the study of the glueball spectrum looks very promising within the RGZ framework. To some extent, our construction can be interpreted as showing a remarkable   consistency of the most recent lattice data for the gluon propagator. These data have in fact been taken as input for the mass parameters entering the RGZ propagator which, in turn, has been employed to achieve a rather good estimate for the masses of the three lightest glueballs.  Also, the present results provide strong evidence that the interplay between the Gribov horizon and the dimension two condensates, as described by the RGZ theory, accounts for a set on nontrivial aspects of the Yang-Mills theory at low energies. The RGZ gluon and ghost propagators turn out to be in agreement with the lattice data in $d=4,3,2$ dimensions. Also, these propagators have allowed us to obtain rather good estimates for the lightest glueballs starting from a first principle calculation of the correlation functions of suitable composite operators.

\section*{Acknowledgements}
D.Dudal and N.Vandersickel are supported by the Research-Foundation Flanders (FWO Vlaanderen).

\end{document}